\documentclass[aip,amsmath,amssymb,reprint]{revtex4-2}

\usepackage{graphicx,bm}
\usepackage[utf8]{inputenc}
\usepackage[T1]{fontenc}
\usepackage{mathptmx,color}

\begin{document}

\title{Dynamics of attractor transitions in Boolean networks under noise}	

\author{Byungjoon Min}
\email[Corresponding author: ]{bmin@cbnu.ac.kr}
 \affiliation{Department of Physics, Chungbuk National University, Cheongju, Chungbuk 28644,  Korea}
 \affiliation{Advanced-Basic-Convergence Research Institute, Chungbuk National University, Cheongju, Chungbuk 28644, Korea}
 \affiliation{Department of Medicine, University of Florida, Gainesville, Florida 32610, USA}
\author{Jeehye Choi} 
\email{choi.jeehye@gmail.com}
 \affiliation{Advanced-Basic-Convergence Research Institute, Chungbuk National University, Cheongju, Chungbuk 28644, Korea}
\author{Reinhard Laubenbacher} 
 \email{reinhard.laubenbacher@medicine.ufl.edu}
 \affiliation{Department of Medicine, University of Florida, Gainesville, Florida 32610, USA}

\date{\today}

\begin{abstract}
Biological systems operate under persistent noise, which can alter system states
and induce transitions between attractors.
Here, we study the attractor dynamics of Boolean networks focusing on the transitions
between attractors induced by noise.
By computing transition probabilities between attractors, we present 
methods at the attractor level to determine dominance, stability, and diversity of attractors, 
and systematically compare local and global noise.
Whereas global noise leads to attractor behavior
dictated primarily by basin sizes, local noise produces 
structured transition patterns
characterized by enhanced stability, non-trivial dominance patterns, 
and broader exploration of the attractor space.
Our work offers insight into the dynamics of attractors,
showing the importance of transition patterns under noise.
\end{abstract}

\maketitle

\begin{quotation}
Biological systems are inherently noisy, with stochastic fluctuations often driving 
transitions between functional states. Despite this, much of the analysis of Boolean 
networks, one of the common theoretical frameworks for gene regulation and other biological 
processes, has overlooked the dynamic impact of noise. 
Motivated by this gap, this study provides a framework to 
address the attractor dynamics in Boolean networks under noise.
By constructing and analyzing transition probabilities among
attractors, we quantify
dominance and stability of attractors, and the diversity of attractor distributions.
We further compare the effects of local state-flip 
noise and global randomization, showing the distinct patterns of attractor
dynamics.
\end{quotation}

\section{Introduction}

Boolean networks have been widely used as mathematical models 
for various biological systems, in particular the study of gene regulatory 
networks \cite{kauffman1969,karlebach2008,shmulevich2002prob}, metabolic 
networks  \cite{areejit2008}, signal transduction networks \cite{helikar2008}, and 
neural networks \cite{palm1996}.
These models provide a simple yet powerful framework to capture the underlying interactions 
and dependencies between components \cite{aldana2003}. 
In the classical Boolean networks, each node represents 
a binary state ($0$ or $1$), and the state of each node is updated based on a Boolean function 
that depends on the states of its neighboring nodes \cite{kauffman1969,kauffman_book,saadatpour2013}.
Many variants of Boolean network models have since been proposed to better reflect biological 
complexity, and a wide range of dynamical behaviors have been extensively 
studied \cite{gardner2000,albert2003,lee2008,ghanbarnejad2011,ebadi2014boolean,kadelka2017,paul2020,min2020,gates2021,kadelka2022,parmer2022}.
One of the key features of Boolean networks is that their state trajectories eventually converge 
to stable configurations (fixed points) or recurring patterns (limit cycles), 
so called ``attractors'' \cite{kauffman1969,derrida1986,klemm2005}.
Attractors play a crucial role in understanding the long-term behavior of the system, as they correspond 
to different functional states in biological systems \cite{kauffman1969,albert2003}.

Noise in biological systems is ubiquitous, arising from intrinsic fluctuations 
in biochemical reactions and external environmental variations \cite{elowitz2002,raser2005,raj2008}.
Noise can randomly alter the states of biological systems, independent to the system’s underlying rules
and thus plays a significant role in shaping dynamics \cite{elowitz2002,raj2008,peixoto2009,villegas2016}.
In Boolean networks, noise can be implemented as random state flips of nodes, 
that are independent of the Boolean update rules \cite{peixoto2009,serra2010,villegas2016,park2023}.
From the perspective of attractor dynamics, such noise can induce significant changes: 
rather than remaining permanently trapped in a single attractor, the system may transition 
between different attractors over time \cite{peixoto2009,serra2010,kuhlman2014}.
As a result, understanding how attractors behave under noise is essential 
for capturing the full dynamical properties of biological systems \cite{kuhlman2014}. It also allows 
for a more complete view of biological stability and variability under noisy environments. 
A related modeling framework is the probabilistic Boolean networks \cite{shmulevich2002prob,shmulevich2002}, 
where state transitions are governed by probabilities rather than deterministic rules, sharing
similar notions of stochastic effects in Boolean networks. 
In contrast to the probabilistic framework, where randomness is incorporated into the update 
rules, the noise considered here operates independently of the system’s deterministic rules.

Given the importance of understanding how noise influences attractor dynamics,
it is crucial to analyze the transition patterns between attractors under noise 
and their long-term dynamics in Boolean networks.
Among existing approaches, Derrida analysis has been widely used to capture 
the spread of perturbations by assessing the average sensitivity 
in Boolean networks \cite{derrida1986,luque1997,shmulevich2004,daniels2018}.
However, this method is limited in its ability to address the the transition patterns of attractors induced by noise, as it does not explicitly include stochastic effects and, moreover, operates at 
the node level rather than the attractor level \cite{park2023}.
To this end, we propose a framework for analyzing attractor dynamics 
based on transition probabilities between attractors under noise.
This framework enables us to systematically assess dynamical structure 
of attractors in Boolean networks and understand how noise shapes 
their long-term behaviors.

The remainder of this paper is organized as follows. In Sec. II, we begin 
by introducing Boolean networks under noise. Next, in Sec. III, we describe 
the mathematical framework for quantifying transition probabilities between 
attractors and for describing the dynamics of Boolean networks on the attractor level.
In Sec. IV, we show how our framework can be used to explore attractor-level 
dynamics in Boolean networks. Our analysis allows us to extract 
several attractor-level quantities, such as the frequencies with which 
attractors are visited, attractor stability, and the diversity of  
attractor distributions. 
We also apply our analysis to real-world Boolean networks to examine 
the applicability our methods to empirical data. 
Finally, in Sec. V, we summarize our results and discuss the implications 
of our study and potential directions for future research.

\begin{figure*}[!]
\includegraphics[width=\linewidth]{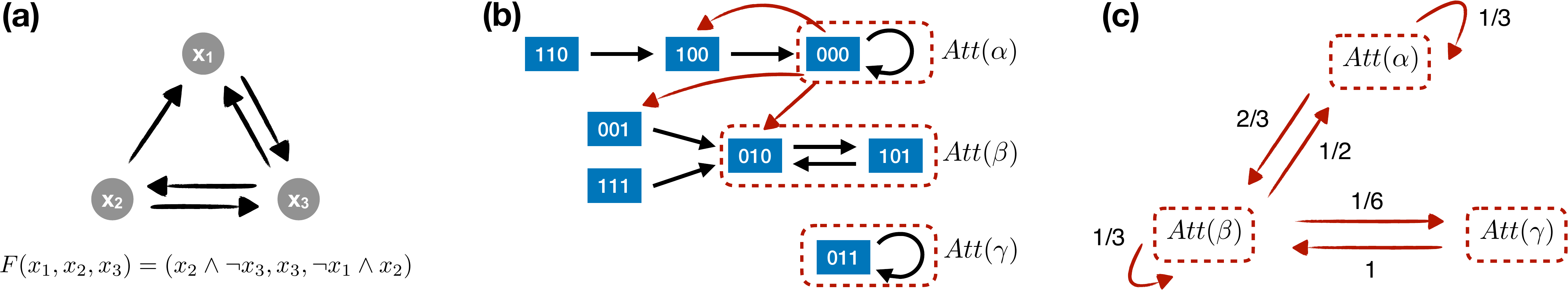}
\caption{
(a) Wiring diagram of a Boolean network and its Boolean function are shown. 
(b) State space of the Boolean network and its attractors are identified. 
A random flip of a node due to local noise results in the transition of the 
state $(0,0,0)$, as illustrated in the diagram. 
(c) The probabilities of remaining within the same attractor or transitioning 
to a different attractor due to local noise are illustrated.
}
\label{fig1}
\end{figure*}

\section{Boolean networks under noise}

We consider a Boolean network which consists of $N$ nodes. 
Each node $i$ has a binary state $x_i \in \{ 0,1 \}$, and 
the state of the system at time $t$ is given by the vector $\mathbf{x}(t) = (x_1(t),x_2(t),\cdots, x_N(t))$. 
The evolution of the system follows a deterministic update rule:
\begin{equation}
\mathbf{x}(t+1) = F(\mathbf{x}(t)),
\end{equation}
where $F$ is a Boolean function that governs the dynamics of nodes
based on the states of its input nodes. 
The Boolean function $F$ is predefined and remains fixed throughout the evolution 
of the system. The system updates all nodes in parallel at each time step. 
Since the dynamics are deterministic, the system eventually converges to a single attractor 
in any finite system. These attractors can be either fixed points where the state remains 
a single state or limit cycles where the system oscillates among a finite set of states \cite{kauffman_book}.

An example of a Boolean network with $N=3$ is depicted in Fig.~\ref{fig1}(a).
The wiring structure shows regulatory influences between nodes through directed edges, with 
Boolean functions specifying how each node updates its state based on its inputs.
The corresponding state space of the network is shown in Fig.~\ref{fig1}(b), where 
each node represents a possible configuration of the system and directed edges indicate 
deterministic evolution between states according to the Boolean update rules.
As shown in Fig.~\ref{fig1}(b), all trajectories eventually lead to one of the attractors, which 
can be either a fixed point or a limit cycle.
As an example, consider the trajectory starting from state 
$(1,1,0)$ in Fig.~\ref{fig1}(b).
The state first moves to $(1,0,0)$ and then settles into the fixed-point attractor $(0,0,0)$.
The non-repeating sequence $(1,1,0) \rightarrow (1,0,0)$ constitutes the transient period before the system reaches its attractor.

We introduce stochastic effects by adding noise to the system in the form of random flipping 
of the state of nodes. 
Under noisy conditions, the system evolves through a combination of deterministic Boolean rules 
and stochastic perturbations.
When noise occurs, the state of affected nodes is randomly altered, leading to
deviations from the purely deterministic trajectory \cite{peixoto2009,serra2004genetic,boldhaus2013,villegas2016}.
Such fluctuations can drive the system to transition between attractors, preventing it from 
remaining in a single attractor indefinitely.
In this study, we assume that noise occurs relatively infrequently
compared to the typical transient time required for the system to relax back to an attractor after a perturbation.
If noise were too frequent, the system would no longer follow the logic of a Boolean network but
instead behave like a randomly fluctuating system, which would lose biological relevance.

Our main objective is to propose a framework for attractor-level
analysis in Boolean networks under noise.
Using this framework, we study various aspects of attractor
dynamics, such as how stable each attractor is to perturbations,
which attractors are more frequently visited in the presence of
noise, and how noise shapes the patterns of transitions
between attractors.

\section{Attractor dynamics via transition probabilities}

In this section, we propose a framework for analyzing attractor dynamics 
under noise by constructing a transition matrix that encodes the probabilities 
of transitions between attractors. Using this matrix, we examine 
the transition patterns of attractors in noisy environments, 
including both local and global noise, and show how it can be 
used to predict key quantities of attractor dynamics.

\subsection{Transition probabilities between attractors}

We first simulate the time evolution of states in a given Boolean network and identify 
all possible attractors. Each state in the network converges to 
one of these attractors after passing through transient states.
We introduce lcoal noise by flipping the value of a randomly chosen node in a given 
attractor state, and simulate the dynamics to determine the resulting attractor.
This process mimics external perturbations that disrupt the deterministic evolution of the system.
Since noise acts intermittently, the Boolean network remains in attractors for 
most of the time rather than in transient states. Therefore, we restrict our analysis 
to attractor states, as the system stays in them with high probability.
By repeating this process for every state in each attractor, we estimate the transition 
probabilities between attractors.

The specific transition probability from attractor $\alpha$ to attractor $\beta$ is 
computed as follows. Let $\ell_\alpha$ be the length of attractor $\alpha$, meaning that 
it consists of $\ell_\alpha$ states. For each state in $\alpha$, we flip each of the $N$ nodes 
once, one flip per node, and count the number of instances $\nu_{\alpha \beta}$ in which 
the system transitions to attractor $\beta$.  
This results in a total of $N \ell_\alpha$ flipping attempts across all states in attractor $\alpha$.  
When $\nu_{\alpha \beta}$ transitions occur from attractor $\alpha$ to attractor $\beta$ out 
of $N \ell_\alpha$ trials, the corresponding transition probability is defined as  
\begin{align}
m_{\alpha\beta} = \frac{\nu_{\alpha\beta}}{N \ell_\alpha}.
\end{align}
Note that the system may remain in the same attractor after a flip, which corresponds to $m_{\alpha\alpha}$.

A schematic illustration of a single-node flip is shown in Fig.~\ref{fig1}(b).  
For example, consider the attractor state $(0,0,0)$.  
In this case, flipping one of the three nodes results in the states $(1,0,0)$, $(0,1,0)$, 
and $(0,0,1)$, each of which occurs with equal probability $1/3$ under noise.
By repeating this procedure for all states in every attractor, we estimate 
the transition probabilities between attractors.  
The resulting attractor-to-attractor transitions are represented as a directed network, shown 
in Fig.~\ref{fig1}(c), where nodes correspond to attractors and edge weights indicate 
the transition probabilities induced by noise.

\subsection{Attractor dynamics from transition matrices}

From the computed transition probabilities, we can construct a transition matrix
\begin{align}
M =
\begin{pmatrix}
m_{11} & m_{12} & \cdots & m_{1n} \\
m_{21} & m_{22} & \cdots & m_{2n} \\
\vdots & \vdots & \ddots & \vdots \\
m_{n1} & m_{n2} & \cdots & m_{nn}
\end{pmatrix}
\end{align}
where each element $m_{\alpha \beta}$ represents the probability of transitioning
from attractor $\alpha$ to attractor $\beta$ under noise. 
The matrix $M$ defines a Markov chain over the set of attractors,
where transitions are governed by perturbations.  
Each element $m_{\alpha \beta}$ represents the probability of transitioning from 
attractor $\alpha$ to attractor $\beta$, satisfying $0 \leq m_{\alpha \beta} \leq 1$ 
and $\sum_{\beta} m_{\alpha \beta} = 1$ for all $\alpha$.

The diagonal elements $m_{\alpha \alpha}$ represent the probability that the system 
remains in the same attractor $\alpha$ after flipping a node.  
We interpret this quantity as a measure of attractor stability; that is, higher values 
indicate more stable attractors, whereas lower values imply a higher tendency 
to transition to other attractors under noise.  
The off-diagonal elements $m_{\alpha \beta}$ with $\alpha \neq \beta$ quantify the 
probabilities of the attractor transitions from $\alpha$ to $\beta$.

The transition matrix $M$ captures the attractor-level dynamics in
the Boolean networks in the presence of noise.
To be specific, let $\vec{u}(t) = (u_1(t), u_2(t), \dots, u_{n}(t))$ 
denote the vector whose component $u_\alpha(t)$ is the probability 
that the system occupies the attractor $\alpha$ at time $t$.
The evolution of these probabilities follows 
\begin{equation}
\vec{u}(t+1) = M^{\mathsf T} \vec{u}(t),
\end{equation}
where $M^{\mathsf T}$ denotes the transpose of the matrix $M$.
Thus, the transition matrix $M$ encodes
the coarse-grained dynamics at the attractor level.

We can also obtain the dominance of attractors through its long-term stationary distribution.  
The transition matrix has a principal eigenvalue $\lambda_1 = 1$ with a corresponding 
eigenvector $\vec{v}$ satisfying $M^{\mathsf T} \vec{v} = \vec{v}$.
The components of $\vec{v}$ form the stationary distribution over attractors, where 
$v_{\alpha}$ represents the fraction of time the system spends in attractor $\alpha$.  
In this sense, attractors with larger $v_{\alpha}$ are considered dominant, as the system 
tends to stay in them more frequently under noise.
In our analysis, we focus only on non-degenerate cases where the principal 
eigenvalue has a unique eigenvector.
In degenerate cases, the stationary distribution is not uniquely defined.
Such degeneracy typically arises when the attractor transition structure 
decomposes into multiple disconnected components, preventing a unique 
value of dominance. For this reason, degenerate cases are excluded 
from our analysis.

\subsection{Global randomization and attractor basins}

For comparison, we consider a global noise scenario  
in which noise simultaneously and randomly alters the states of all nodes.  
In this case, the entire state rather than a single node undergoes a random change.  
Specifically, each node in the Boolean network is independently assigned $0$ or $1$ with 
equal probability when noise is applied.  
We refer to this scenario as global randomization, as flipping is applied to 
the entire network rather than to individual nodes.  
It is important to note that this type of perturbation is fundamentally different 
from the local noise model considered in our study, where only a single node’s state is flipped.
The global randomization erases all memory of the previous state since 
the system is entirely reset to a random point in the state space.

In this setting, the transition matrix is determined solely by the basin sizes of the attractors. 
Here, the ``basin'' of an attractor refers to the set of all initial states that converge to that 
attractor, and its relative size, $b$, reflects how likely a randomly selected initial state converges 
to that attractor. Since global randomization selects states uniformly at random from the full state 
space, the probability $g_{\alpha\beta}$ that global noise drives the system from attractor $\alpha$ 
to attractor $\beta$ depends solely on the basin size $b_{\beta}$ of $\beta$. To be specific,
$g_{\alpha\beta} = b_{\beta}/ N $,
where $b_{\beta}$ denotes the normalized basin size of attractor $\beta$. As a result, all columns are identical, and the eigenvector associated with the principal eigenvalue 
is given by the basin size distribution, $b_\alpha$.

Therefore, under global noise, attractors with larger basins exhibit higher probabilities 
of being reached, and basin size becomes the primary indicator of long-term behavior in this scenario. 
This observation provides a mathematical rationale for the use of basin size 
as larger basins are associated with greater importance of attractors  \cite{grebogi1987,klemm2005,louzada2012monte,krawitz2007basin,paul2020}. 
However, this correspondence does not hold under the local noise model, 
where transitions depend on how individual attractors respond to 
single-node perturbations.

\section{Results}

In this section, we analyze the dynamics of Boolean networks 
under noise by using the attractor-level description.
By representing hopping between attractors by a transition matrix, we can study 
attractor dynamics in a simple and compact manner.
To this end, we generate random regular networks with degree 
$k$ as the underlying structures, and 
assign Boolean functions randomly according to a given bias parameter $p$,  
which represents the probability that the output of a Boolean function is $0$.  
In our analysis, we allow self-regulation.
For comparison across different values of $k$, 
we use the average sensitivity $s = 2p(1-p)k$ \cite{derrida1986,luque1997}
as a unified control parameter. The sensitivity $s$ quantifies the expected number of node 
state changes caused by flipping a single input and serves as a standard measure of the 
system's dynamical sensitivity.
We also apply our analysis to empirical Boolean networks.

\subsection{Dominance and stationary distribution of attractors}

\begin{figure}
\includegraphics[width=\linewidth]{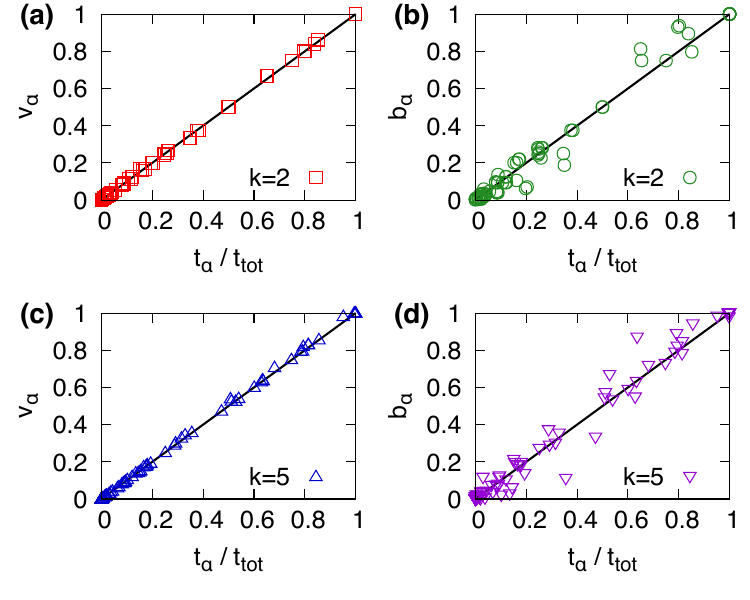}
\caption{
Comparison between the empirically measured time fractions spent 
in attractors, $t_\alpha/t_{\text{tot}}$, and the principal eigenvector components $v_{\alpha}$ 
of the transition matrix is shown for random regular networks with $N = 20$ and $k = 2,5$.
Panels (a,c) show comparisons with eigenvector predictions, while panels (b,d) 
show comparisons with basin sizes.
}
\label{fig4}
\end{figure}

The transition matrix provides an efficient way for identifying the dynamical 
importance of attractors under noise. 
As discussed in Sec.~IIIB, its principal eigenvector $\vec{v}$
offers a theoretical estimate of the fraction of time spent in each attractor.
To validate this prediction, we measure the average duration that the system spends 
in different attractors of Boolean networks under noise. 
For each simulation, we track the transitions between attractors and record 
the time $t_\alpha$ spent in each attractor $\alpha$.  
Specifically, we measure the time fractions $t_\alpha / t_{\text{tot}}$, 
where $t_{\text{tot}}$ is the total simulation time.
In these simulations, the noise rate is set to $0.001$, and 
one time step corresponds to an update of all nodes.

In Fig.~\ref{fig4}, we compare the components $v_\alpha$ of principal eigenvector
with the numerically measured time fractions. Figures~\ref{fig4}(a,c) show that the eigenvector 
accurately predicts the stationary distribution, while Figs.~\ref{fig4}(b,d) 
show the relationship between basin sizes and the empirical time fractions.
The Pearson correlation between $v_\alpha$ and the measured values 
$t_{\alpha}/t_{\text{tot}}$ exceeds 0.99, indicating the system's stationary
distribution is well predicted by the leading eigenvector of the transition matrix.
The small discrepancies arise when noise occurs during transient periods 
before the system fully relaxes back to an attractor.

Additionally, this correspondence implies that attractor dynamics 
can be interpreted as a random walk on the attractor transition 
network, with transitions governed by the probabilities $m_{\alpha\beta}$.  
This is conceptually related to the PageRank algorithm \cite{page} or eigenvector 
centrality \cite{newman_book,bonacich2007,bmin2018}, in which the stationary 
distribution of a Markov chain determines node importance. 
In our framework, attractors play the role of nodes, and their importance 
is reflected in the time that the system spends in each under noise 
predicted theoretically by the leading eigenvector. 
Thus, this approach provides an efficient way to identify dominant 
attractors and to predict the asymptotic behavior of Boolean networks 
under stochastic perturbations.

\subsection{Stability of attractors under noise}

\begin{figure}
\includegraphics[width=\linewidth]{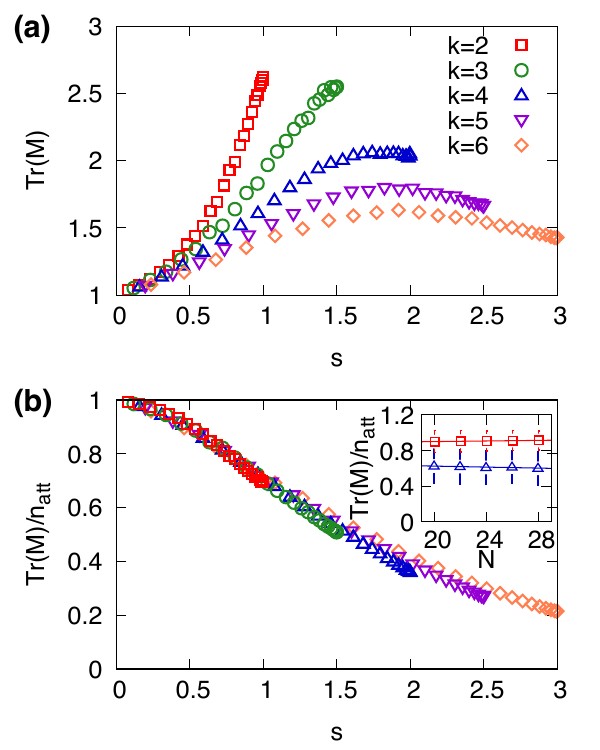}
\caption{(a) The trace $\mathrm{Tr}(M)$ of transition matrix for random Boolean networks on random
regular networks with size $N=20$ and various degrees $k$ with respect to the sensitivity $s$ is shown.
(b) The trace normalized by the number of attractors, $\mathrm{Tr}(M)/n_{\mathrm{att}}$ is shown. 
Inset shows $\mathrm{Tr}(M)/n_{\mathrm{att}}$ as a function of size $N$
for $(k,s)=(2,0.75)$ for squares and $(k,s)=(4,1.5)$ for triangles.
}
\label{fig2}
\end{figure}

An important concept in Boolean networks is the stability of attractors, which refers to the system’s ability to return to the same attractor after a perturbation.
The stability of an attractor $\alpha$ can be directly quantified by the diagonal 
element $m_{\alpha \alpha}$ of the transition matrix.
To evaluate stability at the network level, we  compute the trace of 
the transition matrix, $\mathrm{Tr}(M)$.
A larger trace indicates that attractors in Boolean networks
are more stable to noise. 
For local noise, the trace can be computed from the transition matrix 
constructed by explicitly evaluating the effects of all local perturbations.
For the global randomization 
in which the system's state is reset entirely at random, 
each term $m_{\alpha \alpha}$ becomes $b_{\alpha}/N$ where 
$b_\alpha$ is the basin size.
Thus, the trace of the transition matrix equals one by definition, 
as it results from the normalization of basin sizes. 
Thus, we obtain $\mathrm{Tr}(M) = 1$ for the global noise.

We compare the stability of attractors for local and global noise in Fig.~\ref{fig2}(a).
The trace $\mathrm{Tr}(M)$ for local noise is consistently 
greater than unity across various values of the connectivity $k$ and sensitivity $s$,
on random regular networks with $N=20$. 
These observations suggest that Boolean networks have an intrinsic resilience to local 
perturbations, a feature that has been recognized since the pioneering work in 
Boolean networks \cite{kauffman1969}.
Our findings also suggest the importance of assessing stability through 
$\mathrm{Tr}(M)$, which directly reflect the system's resilience to noise.

In Fig.~\ref{fig2}(b), we show the average attractor stability under local noise, defined 
as $\mathrm{Tr}(M)/n_{\mathrm{att}}$, where $n_{\mathrm{att}}$ is the number of attractors 
in a given network. This normalization is introduced because the number of attractors 
varies across different network instances and parameter settings. The results show 
that average attractor stability tends to decrease as the average sensitivity $s$ increases.
The inset of Fig.~\ref{fig2}(b) shows the relation between $\mathrm{Tr}(M)/n_{\mathrm{att}}$
and the network size $N$, which appears nearly constant within the range studied. 
However, because the tested sizes are small, the nature of this size dependence 
requires further investigation. This behavior is also different from 
the global-noise case, where the attractor stability becomes
$1/n_{\mathrm{att}}$ and is independent of $s$ and $k$.

In the case of local noise, we can also estimate a lower bound for
attractor stability, which depends on the network parameters $s$ and $k$.  
Within the annealed approximation \cite{derrida1986}, we can estimate
the one-step probability $P_{\mathrm{ret}}^1$ that a single-node perturbation 
disappears after one update step, i.e., that the system returns 
exactly to the pre-perturbation state, as
\begin{align}
P_{\mathrm{ret}}^{(1)} \approx \left( 1 - \frac{s}{k} \right)^{k}.
\end{align}
This quantity becomes a lower bound on the attractor stability, because perturbed 
states may still belong to the same attractor basin even when they
do not immediately return to the original state.
This expression captures the dependence of the stability on the parameters $s$ and $k$, and it also shows that the bound decreases monotonically as the sensitivity $s$ increases.


\begin{figure*}
\includegraphics[width=\linewidth]{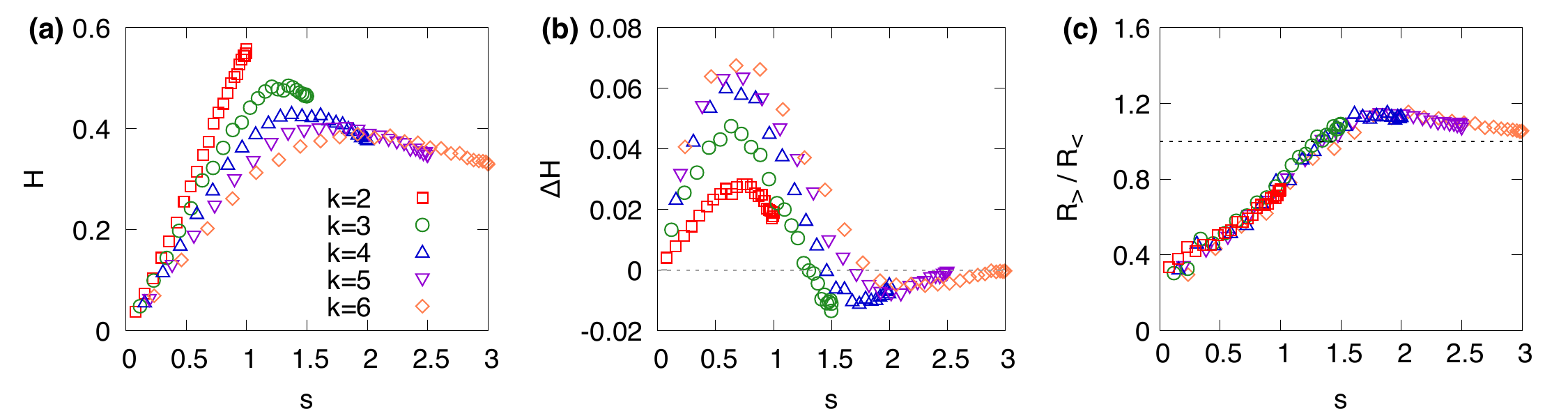}
\caption{
(a) Entropy $H$ of the principal eigenvector as a function of network sensitivity $s$,
(b) the difference $\Delta H$ between entropy from basin sizes and from the eigenvector distribution, and
(c) the ratio $R_>/R_<$ comparing transition bias toward larger versus smaller basins.
All results are obtained on random regular networks with $N = 20$ and various values of $k$.
}
\label{fig5}
\end{figure*}

\subsection{Entropy of attractor distributions}

We further study how broadly the system explores the attractor space.
To this end, we compute the normalized Shannon entropy $H$ of the principal 
eigenvector $\vec{v}$ of the transition matrix, defined as
\begin{align}
H = - \frac{1}{H_{\max}} \sum_\alpha v_\alpha \log_2 v_\alpha,
\end{align}
where $v_\alpha$ represents the steady-state probability that the system occupies 
attractor $\alpha$ under noise, and $H_{\max} = \log_2 n_{\text{att}}$ 
is the maximum entropy possible for a given number of attractors.
The normalization ensures that $H \in [0,1]$.
High entropy indicates that the system frequently visits many attractors with 
similar probabilities, reflecting broad exploration and dynamical diversity.
In contrast, low entropy implies strong localization around a few dominant attractors.
Figure~5(a) shows that the Shannon entropy $H$ as a function of $s$ for various $k$.

To assess how this behavior compares with the global randomization,
we compute the entropy difference $\Delta H$ between the stationary distribution 
and the basin size distribution:
\begin{align}
\Delta H = -\frac{1}{H_{\max}} \sum_\alpha \left(v_\alpha \log_2 v_\alpha- b_\alpha \log_2 b_\alpha \right),
\end{align}
where $b_\alpha$ is the normalized basin size of attractor $\alpha$.
Positive values of $\Delta H$ indicate that the system explores a broader subset 
of attractors under local stochastic dynamics than would be expected based on 
the global randomization.
Figure~5(b) shows that $\Delta H$ is positive at low sensitivity, 
showing that attractor dynamics become more spread out than what 
the global randomization would predict.
As $s$ increases, $\Delta H$ gradually decreases and hovers near 
zero above $s>1.5$, suggesting convergence toward the global randomization.

We examine transition 
asymmetries between attractors. Figure~5(c) shows the ratio of 
transition probabilities between attractors of different 
basin sizes. We compute the ratio $\frac{m_{\alpha \beta}}{b_\alpha}$, 
where $m_{\alpha \beta}$ is the probability of transitioning from 
attractor $\alpha$ to $\beta$, and $b_\alpha$ is the 
basin size of the source attractor. We then compare this value 
for transitions into larger basins ($R_{>}$) versus smaller ones ($R_{<}$), 
using the ratio $R_{>}/R_{<}$ to quantify the directionality of transitions. 
Our results indicate that for $s < 1.5$, transitions 
tend to favor smaller basins, i.e., $R_{>}/R_{<} < 1$.
This pattern of transitions enhances the occupation of small-basin 
attractors relative to their basin sizes.
At higher sensitivities, this tendency diminishes and the ratio approaches $1$, 
implying that the transition dynamics become more random, 
consistent with global randomization.

\begin{figure}
\includegraphics[width=\linewidth]{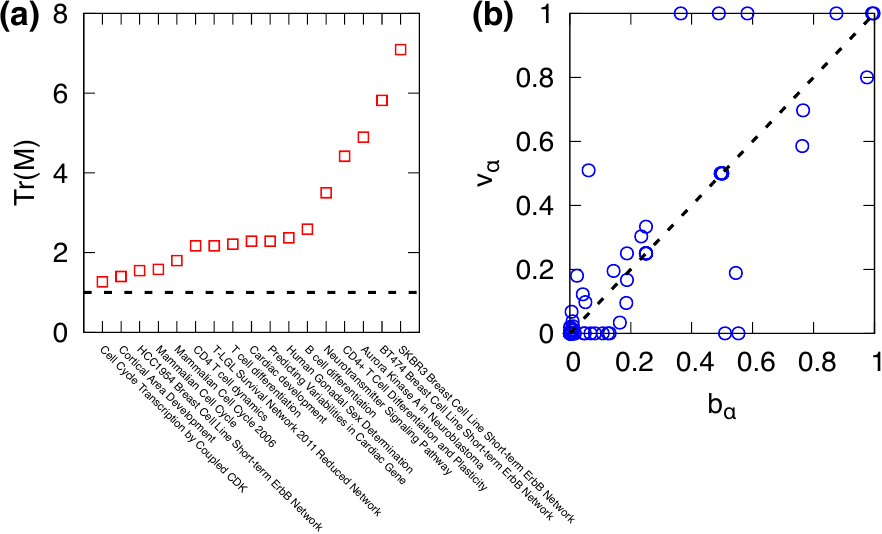}
\caption{
(a) The trace of the transition matrix, $\mathrm{Tr}(M)$, is shown
in descending order for real-world Boolean networks.
(b) A comparison between the principal eigenvector components $v_{\alpha}$ and 
basin sizes $b_{\alpha}$ for all attractors across the real-world networks is presented.
}
\label{fig6}
\end{figure}

\subsection{Application to real-world Boolean networks}

Finally, we apply our framework to real-world Boolean networks obtained from 
the Cell Collective database \cite{helikar2012,cellcollective}. These datasets include biological systems 
such as gene regulatory networks, signaling pathways, and developmental processes. Basic properties of the real-world Boolean networks used in this section are presented in the Table~1. in the Appendix.
We found that the trace of the transition matrix, $\mathrm{Tr}(M)$, exceeded unity 
consistently across all tested networks as shown in Fig.~\ref{fig6}(a). 
This indicates that, similar to random Boolean networks, real-world Boolean networks 
exhibit enhanced attractor stability under local noise.

We further compared the stationary distributions predicted by the principal eigenvectors 
with the basin size distributions. In Fig.~\ref{fig6}(b), each data point corresponds to 
an individual attractor from 17 different real-world Boolean networks, where the horizontal 
axis represents the normalized basin size and the vertical axis indicates the corresponding 
component of the principal eigenvector. We found that basin sizes deviate from 
eigenvector, showing that local and global noise produce distinct 
patterns of attractor dynamics.
Some extreme data points with larger basin size and $v=0$ correspond to unused 
attractors in noisy environments that still possess relatively large basin sizes.
These unusual attractors show interesting anomalies and become a 
potentially direction for future research.

\section{Discussion}

In this study, we proposed a framework for analyzing attractor dynamics in 
Boolean networks under noise, based on a representation 
of transitions between attractors.
This approach offers insight into key properties of attractor behavior 
such as long-term dominance and stability.
We also compare the local and global noise, showing that they
lead to different behaviors in attractor dynamics. 
To be specific, local noise shows structured and non-trivial
transition patterns, whereas global noise yields attractor 
dynamics largely determined by basin sizes.

Further applications of our method to real-world Boolean networks are 
needed to deepen our understanding of attractor dynamics under noise.
It remains important to assess how generally these results hold, especially in large-scale networks where one can use sampling-based or approximate methods, which represents an important direction for future work.
In addition, future research could explore how specific topological features, 
such as modularity \cite{kadelka2022,wheeler2024}, 
redundancy \cite{gates2021,kadelka2024}, and/or 
the presence of canalizing functions 
influence attractor transitions \cite{moreira2005,kadelka2017,paul2020}.
Extending the framework to include asynchronous updates \cite{ghanbarnejad2011}, 
multi-node perturbations \cite{boldhaus2013}, or correlated noise \cite{ai2003} may 
provide further insight into biological function.

\begin{acknowledgments}
This research was supported in part by the National
Research Foundation of Korea (NRF) grant funded by
the Korea government (MSIT) (No. 2020R1I1A3068803),
by Global - Learning \& Academic research institution 
for Master’s $\cdot$ PhD students, and Postdocs (LAMP)
Program of the National Research Foundation of Korea
(NRF) grant funded by the Ministry of Education (No.
RS-2024-00445180) and
by the IITP(Institute of Information \& Coummunications Technology Planning \& Evaluation)-ITRC(Information Technology Research Center) grant funded by the Korea government(Ministry of Science and ICT)(IITP-2025-RS-2024-00437284).
This work was conducted during the research year of Chungbuk National University in 2024-2025.
The research of R.L. was partially supported by 
the grants NSF DMS-2424635, NIH R01 AI135128, and NIH R01 HL169974-01.
\end{acknowledgments}

\appendix
\section{Characteristics of real-world Boolean networks}

In this appendix, we summarize the properties of a set of real-world Boolean networks used in our
analysis, obtained from the Cell Collective database \cite{cellcollective}. 
For each network, we report its number of nodes, mean in-degree, and mean bias of its Boolean functions in Table.~1.

\begin{table}
\centering
\caption{Real-world Boolean networks}
\resizebox{\columnwidth}{!}{
\begin{tabular}{cccc}
\hline
\textbf{Boolean networks} & \textbf{bias} & \textbf{degree} & \textbf{size} \\
\hline
Cortical Area Development & 0.163 & 2.8 & 5 \\
Cell Cycle Transcription by Coupled CDK  & 0.306 & 2.111 & 9 \\
Mammalian Cell Cycle 2006 & 0.270 & 3.5 & 10 \\
Predicting Variabilities in Cardiac Gene & 0.303 & 2.533 & 15 \\
Cardiac development & 0.303 & 2.533 & 15 \\
CD4 T cell dynamics – Workshop & 0.079 & 1.875 & 16 \\
Neurotransmitter Signaling Pathway & 0.414 & 1.375 & 16 \\
SKBR3 Breast Cell Line Short-term ErbB Network & 0.084 & 2.563 & 16 \\
BT474 Breast Cell Line Short-term ErbB Network & 0.075 & 2.875 & 16 \\
HCC1954 Breast Cell Line Short-term ErbB Network & 0.087 & 2.875 & 16 \\
CD4+ T Cell Differentiation and Plasticity & 0.065 & 4.333 & 18 \\
T-LGL Survival Network 2011 Reduced Network & 0.25 & 2.389 & 18 \\
Human Gonadal Sex Determination & 0.289 & 4.158 & 19 \\
Mammalian Cell Cycle & 0.2422 & 2.55 & 20 \\
B cell differentiation & 0.296 & 1.773 & 22 \\
Aurora Kinase A in Neuroblastoma & 0.309 & 1.869 & 23 \\
T cell differentiation & 0.325 & 1.478 & 23 \\
\hline
\end{tabular}}
\end{table}

\bibliography{boolean}

\end{document}